\documentclass[10pt,thmsa]{article}
\input epsf
\begin{document}

\title{The Sunyaev-Zel'dovich effect revisited }
\author{A. Sandoval-Villalbazo$^a$ and L.S.
Garc\'{\i}a-Col\'{\i}n$^{b,\,c}$ \\$^a$ Departamento de Ciencias,
Universidad Iberoamericana \\Lomas  de Santa Fe 01210 M\'{e}xico
D.F., M\'{e}xico \\E-Mail: alfredo.sandoval@uia.mx \\$^b$
Departamento de F\'{\i}sica, Universidad Aut\'{o}noma Metropolitana \\
M\'{e}xico D.F., 09340 M\'{e}xico
\\$^c$ El Colegio Nacional, Centro Hist\'{o}rico 06020 \\
M\'{e}xico D.F., M\'{e}xico \\E-Mail: lgcs@xanum.uam.mx}

\maketitle

\begin{abstract}
The well known Sunyaev-Zel'dovich (SZ) effect is reexamined using
a Doppler shift type mechanism arising from the scattering of
photons by electrons in an optically thin gas. The results are in
excellent agreement with the observational data as well as with
the results obtained using the diffusive pictures. A comparison of
the results here obtained with other approaches is thoroughly
discussed, as well as some important extensions of this method to
other aspects of the SZ effect.
\end{abstract}

\section{\textbf{Introduction}}

As it is well known today, the cosmic microwave background
radiation (CMBR) that fills the universe exhibits an almost
perfect blackbody Planckian spectrum corresponding to a
temperature of 2.726 K. A little over thirty years ago, it was
predicted by the astrophysicists R.A Sunyaev and Ya. B. Zel'dovich
that this spectrum could be distorted when photons from this
radiation would penetrate large structures such as the hot
intracluster gas now known to exist in the universe~\cite{SZ1}.
This distortion would arise from the interaction between the
photons and electrons which constitute the plasma responsible for
an important main contribution to the total mass contained in the
cluster. Such distortion is only a very small effect changing the
brightness of the spectrum by a figure of the order of 0.1
percent. This effect is now called the Sunyaev-Zel'dovich effect
and its detection is at present a relatively feasible task due to
the modern observational techniques available. Its main interest
lies on the fact that it provides information to determine
important cosmological parameters such as Hubble's constant and
the baryonic density. Since all these and other important facts
are readily available in the somewhat broad literature on the
subject~\cite{SZ1}-\cite{Steen1}, we will not pursue any further
details of its cosmological implications.

In this work we are mainly concerned with the nature of the
various interpretations that have been provided to explain the
effect and to propose another one which we believe, is much
simpler to grasp than the others, and offers a much more direct
way to account for the relativistic corrections, as well as the
influence of $z$, the redshift factor, in the equations for the
distorted spectrum.

The main idea behind the method we want to discuss is that photons
incident on the hot electron gas of the plasma change their
frequency by an absortion-emission process, so that the intensity
of the spectral line is changed through the Doppler effect.

The overall effect is due to those photons that happen to be
captured by an electron, an event which is in general unlikely to
occur specially in an optically thin gas. Thus, an electron moving
with a given thermal velocity emits (scatters) a photon with a
certain incoming frequency $\nu _o$ and outgoing frequency $\nu $.
The line breath of this process is readily calculated from
elementary kinetic theory taking into account that the media in
which the process takes place has an optical depth directly
related to the usual Compton parameter $y$. When the resulting
expression is convoluted with the incoming flux of photons
obtained from Planck's distribution, one easily obtains the
disturbed spectra. The results derived from the final equation are
shown to be in agreement with those obtained from the
observational data. To present these results we have divided the
paper as follows. In section II a brief review of the previous
methods developed to explain the SZ effect is given. This will
allow a thorough comparison with our proposal. In Section III we
develop our ideas and give the comparison between the mathematical
results with both, those derived before and the observational
ones. Section IV is left for some concluding remarks and future
directions of this work.

\section{The Diffusion-Scattering  Picture of the SZ effect.}

As it was clearly emphasized by the authors of this discovery in
their early publications ~\cite{SZ1}~\cite{SZ2} as well as by
other authors, the distortion in the CMBR spectrum by the
interaction of the photons with the electrons in the hot plasma
filling the intergalactic space is due to the diffusion of the
photons in the plasma which, when colliding with the isotropic
distribution of a non-relativistic electron gas, generates a
random walk. The kinetic equation used to describe this process
was one first derived by Kompaneets back in 1956
~\cite{Komp}-\cite{Weymann}. For the particular case of interest
here, when the electron temperature $Te\sim 10^8K$ is much larger
than that of the radiation, $T\equiv T_{Rad}\;(2.726\,\,K)$, the
kinetic equation reads ~\cite{Peebles}
\begin{equation}
\frac{dN}{dy}=\nu ^2\frac{d^2N}{d\nu ^2}+4\nu \frac{dN}{d\nu }  \label{Uno}
\end{equation}

Here $N$ is the Bose factor, $N=(e^x-1)^{-1}$ , $x=\frac{h\nu
}{kT}$, $\nu $ is the frequency, $h$ is Planck's constant, $k$
Boltzmann's constant and $y$ the ``Compton parameter'' given by,

\begin{equation}
y=\frac{k\,Te}{m_ec^2}\int \sigma _T\,n_ec\,dt=\frac{k\,Te}{m_ec^2}\tau
\label{dos}
\end{equation}
$m_ e$ being the electron mass, $c$ the velocity of light, $\sigma
_T$ the Thomson's scattering cross section and $n_e$ the electron
number density. The integral in Eq. (\ref{dos}) is usually
referred to as the ``optical depth, $ \tau $'', measuring
essentially how far in the plasma can a photon travel before being
captured (scattered) by an electron. Without stressing the
important consequences of Eq. (\ref{Uno}) readily available in
many books and articles ~\cite{Peebles}-\cite{imp1}, we only want
to state here, for the sake of future comparison, that since the
observed value of $N_o(\nu )$ is almost the same as its
equilibrium value $N_{eq}(\nu )$, to a first approximation one can
easily show that
\begin{equation}
\frac{\delta N}N\equiv \frac{N_o(\nu )-N_{eq}(\nu )}{N_{eq}(\nu )}=y\left[
\frac{x^2e^x(e^x+1)}{(e^x-1)^2}-\frac{4xe^x}{e^x-1}\right]   \label{tres}
\end{equation}
implying that, in the Rayleigh-Jeans region $\left( x<<1\right) $,
\begin{equation}
\frac{\delta N}N\cong -2y  \label{cuatro}
\end{equation}
and in the Wien limit $\left( x\ >>1\right) $ :
\begin{equation}
\frac{\delta N}N\cong x^2y  \label{cinco}
\end{equation}
If we now call $I_o(\nu )$ the corresponding radiation flux for
frequency $ \nu $, defined as
\begin{equation}
I_o(\nu )=\frac c{4\pi }U_v(T)  \label{seis}
\end{equation}
where $U_v(T)=\frac{8\pi \,h\nu ^3}{c^3}N_{eq}(\nu )$ is the energy density
for frequency $\nu $ and temperature $T$ , and noticing that
\begin{equation}
\frac{\delta T}T=\left( \frac{\partial (\ln I(\nu ))}{\partial
(\ln T)} \right) \left( \frac{\delta I}I\right)   \label{siete}
\end{equation}
we have that the change in the background brightness temperature is given,
in the two limits, by
\begin{equation}
\frac{\delta T}T\cong -2y\;,\;x<<1  \label{ocho}
\end{equation}
and
\begin{equation}
\frac{\delta T}T\cong x\,y\;,\;x>>1  \label{nueve}
\end{equation}
showing a decrease in the low frequency limit and an increase in
the high frequency one. Finally, we remind the reader that the
curves extracted from Eq. (\ref{tres}) for reasonable values of
the parameter $``y"$ show a very good agreement with the
observational data.

Nevertheless, many authors, including Sunyaev and Zel'dovich
themselves, were very reluctant in accepting a diffusive mechanism
as the underlying phenomena responsible for the spectrum
distortion. The difficulties of using diffusion mechanisms to
study the migration of photons in turbid media, specially thin
media, have been thoroughly underlined in the literature
~\cite{x}-\cite{y}-\cite{z}. Several alternatives were discussed
in a review article in 1980 ~\cite{imp1} and a model was set forth
by Sunyaev in the same year ~\cite{imp2} based on the idea that
Compton scattering between photons and electrons induce a change
in their frequency through the Doppler effect. Why this line of
thought has not been pursued, or at least, not widely recognized,
is hard to understand. Two years ago, one of us (ASV) ~\cite{yo}
reconsidered the single scattering approach to study the SZ
effect. The central idea in that paper is that in a dilute gas,
the scattering law is given by what in statistical physics is
known as the \emph{ dynamic structure factor}, denoted by $S(k,\nu
)$ where $k=\frac{2\pi } \lambda $, and $\lambda $ is the
wavelength. In such a system, this turns out to be proportional to
$\exp (-\frac{\nu ^2}{w^2})$ where $w$ is the broadening of the
spectral line given by
\begin{equation}
w=\frac{2}c(\frac{2kTe}m)^{1/2}\nu   \label{diez}
\end{equation}
If one computes the distorted spectrum through the convolution integral
\begin{equation}
I\left( \nu \right) =\int_0^\infty I_o\left( \bar{\nu}\right)
S(k,\bar{\nu} -(1-ay)\nu )\,d\bar{\nu}  \label{once}
\end{equation}
where the corresponding frequency shift $\frac{\delta \nu }\nu $
has been introduced through Eqs. (\ref{ocho}-\ref{nueve})
$(a=-2+x)$, one gets a good agreement with the observational data.
This is exhibited in Fig. (1) for  $ y=10^{-5}$. This result is
interesting from, at least, two facts. One, that such a simple
procedure is in agreement with the diffusive picture. This poses
interesting mathematical questions which will be analyzed
elsewhere, specially since the exact solution to Eq. (1) is known
(see Eq. (A-8) ref.~\cite{imp1}). The other one arises from the
fact that this is what triggered the idea of reanalyzing the SZ
effect using elementary arguments of statistical mechanics and
constitutes the core of this paper to be presented next.

\begin{figure}
\epsfxsize=3.4in \epsfysize=2.6in \epsffile{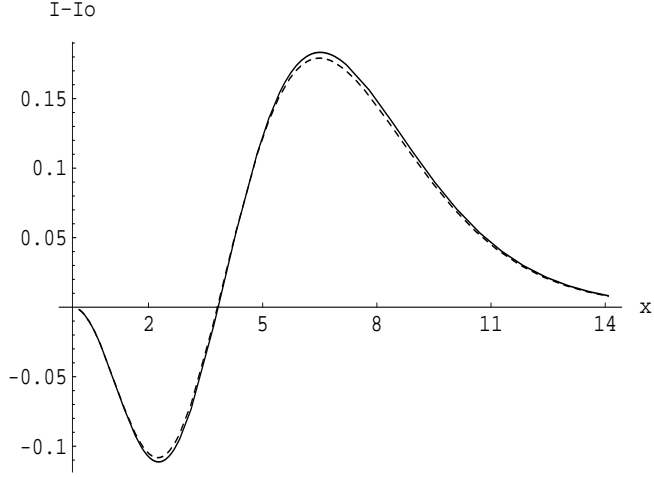}\vspace{0.5cm}
\caption{Comparison of the CMBR SZ distortion as computed by Eqs.
(3,6) (solid line) and Eq. (11) (dashed line) i.e using the
approach of Ref. [13] , with $y=10^{-5}$ and $T=10^{8} K$. $\delta
I$ is measured in $erg$ $s^{-1}$ $cm^{-2}$ $ster^{-1}$, notice
that the figure is scaled by a factor of $10^{18}$.}
\end{figure}
\vspace{0.5cm}

\section{The Doppler Effect Approach.}

We begin this section by simply reminding the reader that if an
atom in an  ideal gas moving say with speed $u_x$ in the $x$
direction emits light of frequency $\nu _o$ at some initial speed
$u_x(0)$, the intensity of the spectral line $I\left( \nu \right)
$ is given by
\begin{equation}
\frac{I\left( \nu \right) }{I_o\left( \nu \right) }=\exp \left[
-\frac{m\,c^2 }{2kT}\left( \frac{\nu _o-\nu }\nu \right) ^2\right]
\label{doce}
\end{equation}

Eq. (\ref{doce}) follows directly from the fact the velocity
distribution function an ideal gas is Maxwellian and that $u_x$,
$u_x(0)$ and $\nu $ are related through the Doppler effect. For
the case of a beam of photons of intensity $I_o\left( \nu \right)
$ incident on a hot electron gas regarded as an ideal gas in
equilibrium at a temperature $T_e$ the full distorted spectrum may
be computed from the convolution integral given by
\begin{equation}
I\left( \nu \right) =\frac 1{\sqrt{\pi }W\left( \nu \right)
}\int_0^\infty I_o\left( \bar{\nu}\right) \exp \left[ -\left(
\frac{\bar{\nu}-f(y)\,\nu }{ W\left( \nu \right) }\right)
^2\right] \,d\bar{\nu}  \label{trece}
\end{equation}
which defines the joint probability of finding an electron
scattering a photon with incoming frequency $\bar{\nu}$\thinspace
and outgoing frequency $ \nu $ multiplied by the total number of
incoming photons with frequency $ \bar{\nu}$. $\frac 1{\sqrt{\pi
}}W\left( \nu \right) $ is the normalizing factor of the Gaussian
function for $f(y)=1$, $ W\left( \nu \right) $ is the width of the
spectral line at frequency $\nu $ and its squared value follows
from Eq.(\ref{diez})
\begin{equation}
W^2\left( \nu \right) =\frac{4kT_e}{m_ec^2}\tau \,\nu ^2=4\,y\,v^2
\label{catorce}
\end{equation}
where $\tau $ is the optical depth whose presence in Eq.
(\ref{catorce}) will be discussed later. The function $f(y)$
multiplying $\nu $ in Eq. (\ref {trece}) is given by $f(y)=1+ay$
where $a=-2$ in the Rayleigh-Jeans limit and $a=xy$ in the Wien's
limit, according to Eqs. (8-9) and the fact that $\frac{ \Delta
\nu }\nu =\frac{\Delta T}T$ for photons. Eq.(\ref{trece}) is the
central object of this paper so it deserves a rather detailed
examination. In the first place it is worth noticing that $I_o(\nu
)$, the incoming flux,is defined in  Eq. (\ref{seis}).  Secondly,
it is important to examine the behavior of the full distorted
spectrum in both the short and high frequency limits. In the low
frequency limit, the Rayleigh-Jeans limit $I_o(\nu )$
$=\frac{2kT\nu ^2}{c^2}$,  so that performing the integration with
$a=-2$ and noticing that $y$ is a very small number, one arrives
at the result
\begin{equation}
\frac{\delta I}I\equiv \frac{I(\nu )-I_o(\nu )}{I_o(\nu )}=-2y,\;x<<1
\label{quince}
\end{equation}
(\ref{quince}) is in complete agreement with the value obtained
using Eq.( \ref{tres}), the photon diffusion equation. In the high
frequency limit where $a=xy$ and $I_o(\nu )$ $=\frac{2h\nu
^3}{c^2}e^{-x}$, a slightly more tedious sequence of integrations
leads also to a result at grips with the diffusion equation,
namely
\begin{equation}
\frac{\delta I}I=x^2y,\;x>>1  \label{dieciseis}
\end{equation}
Why both asymptotic results, the ones obtained with the diffusion
equation and those obtained from Eq. (\ref{trece}) agree so well,
still puzzles us. At this moment we will simply think of them as a
mathematical coincidence. Nevertheless, it should be stressed that
in his 1980 paper, Sunyaev ~\cite{imp2}  reached rather similar
conclusions although with a much more sophisticated method, and
less numerical accuracy for the full distortion curves. The
distorted spectrum for the CMBR radiation may be easily obtained
by numerical integration of Eq. (\ref{trece}) once the optical
depth is fixed, $y$ is determined through Eq.(\ref{dos}) and
$a=-2+x$. The intergalactic gas cloud in clusters of galaxies has
an optical depth $\tau \sim 10^{-2}$ ~\cite{rel}. To compare our
results with the curves obtained from the traditional approach, we
have plotted $\delta I\left( \nu \right) $  for several
temperatures, typical of the hot intra-cluster gas, in the
non-relativistic range, in Figs. (2) and (3).

\begin{figure}
\epsfxsize=3.4in \epsfysize=2.6in \epsffile{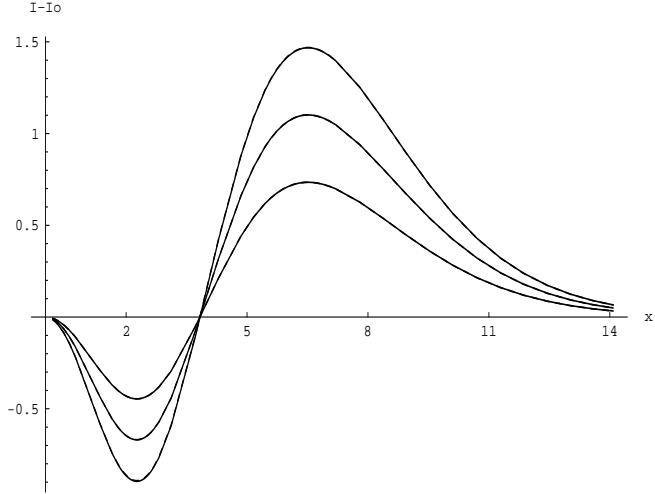} \vspace{0.5cm}
\caption{Comparison of the CMB SZ distortion as computed by Eqs.
(3-6) (dashed line) and Eq. (13) (solid line), with $\tau =
10^{-2}$ , $T=2 KeV$ (lower curves), $T=3 KeV$ (middle curves) and
$T=4 KeV$ (upper curves). Here $a=-2$ in Eq. (13). Since these are
actually six curves, it is clear that the accuracy achieved by Eq.
(13) is remarkable. $\delta I$ is measured in $erg$ $s^{-1}$
$cm^{-2}$ $ster^{-1}$, the figure is again scaled by a factor of
$10^{18}$. }
\end{figure}
\vspace{0.5cm}

\begin{figure}
\epsfxsize=3.4in \epsfysize=2.6in \epsffile{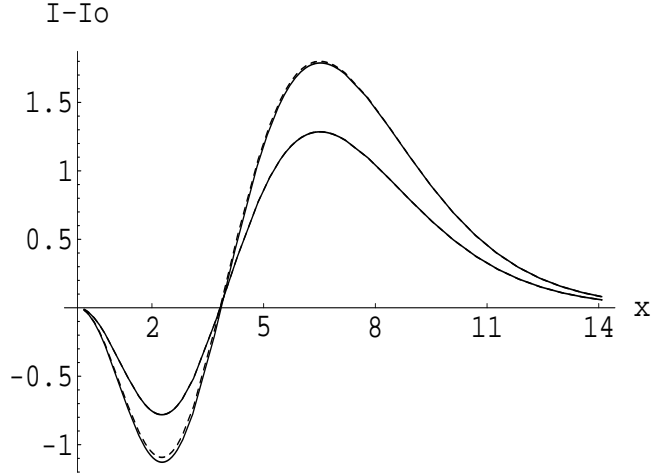} \vspace{0.5cm}
\caption{The same as in figure 2, but taking $\tau = 7 \times
10^{-3}$ . The respective temperatures are $5$,  and  $7~KeV$.}
\end{figure}
\vspace{0.5cm}

From the results obtained one appreciates the rather encouraging
agreement between the observational data and the theoretical
results obtained with the three methods, the diffusion equation,
the structure factor or scattering law approach, and the Doppler
effect. This, in our opinion is rather rewarding and some efforts
are in progress to prove the mathematical equivalence of the three
approaches. From the physical point of view, and for reasons
already given by many authors, we believe that the scattering
Doppler effect picture does correspond more with reality,
specially for reasons that will become clear in the last section.

\section{Discussion of the results}

Eq. (\ref{trece}), the main result of this work hardly needs a
more detailed explanation, it is a direct result of elementary
statistical mechanics, except of course for the factor $\tau $
which appears in the broadening of the spectral line. That this
broadening must somehow depend on $\tau $ is clear. As argued by
many authors in the case of clusters of galaxies, the plasma is
optically thin, so most of the photons are not scattered, and how
many of them are must depend on the optical depth. Now, why in
particular Eq. (\ref{catorce}) is valid remains to be rigorously
shown. In our case it came as a mere accident since in ref.
\cite{yo} the mass taken  in $w$ to perform the calculations was
that of a proton, which turns out to be of the same order of
magnitude as $\frac{m_e}\tau $ for $y=10^{-5}$. There is in fact
one way of understanding this puzzle. In a pure diffusive
(Fickian) process described by a Gaussian function one knows that
the squared value of a single line  width grows as $t$. Looking at
the original equation of Kompaneets and thinking only in its
diffusive terms, one sees that the variable time is replaced by
the Compton parameter "$y$", so that one should expect that, to a
first approximation, the width of the curve grows as $\sqrt{y}$.
Hence, inside the gas, the effective frequency at which the photon
propagates through the gas is, according to Eq. (\ref{catorce}),
proportional to $\sqrt{y} \times \nu $. This argument, which is
based on ideas pertinent to the diffusion mechanism must be
extended to the Doppler picture, but this has not yet been
accomplished.

The two main advantages of Eq. (\ref{trece}) are, firstly, that
its generalization to the relativistic case is straightforward and
indeed, since in relativity theory  one has both the transverse
and parallel Doppler effects~\cite{Moller}, one can easily  study
what their influence is, if any, in the full distorted spectrum.
For the parallel effect the calculations carried out so far agrees
reasonably well with those reported by Rephaeli  and other authors
~\cite{rel} and will be the subject of a forthcoming publication.
The case of the transverse effect is under study. Secondly, in all
the work we have reported here, the cluster of galaxies, the
source of the scattering processes of the CMBR and the electrons,
has been considered a system at rest. One can now extend the
calculations assuming that the cluster recedes to infinity
according to Hubble's law and introduce the redshift factor into
the formalism. The resulting prediction could be compared with the
observational data and extract valuable information on such an
important cosmological parameter. Work in this direction is also
in progress. Concluding, we have the strong conviction that the
Doppler effect approach to study the thermal SZ effect is not only
very clear from the physical point of view, it also contains
ingredients which are promising for further studies more
advantageous than those offered by the diffusive model.

\end{document}